\documentclass[%
 reprint,
nofootinbib,
 amsmath,amssymb,
 aps,
]{revtex4-1}

\usepackage{graphicx}
\usepackage{dcolumn}
\usepackage{bm}

\usepackage[usenames,dvipsnames]{xcolor}

\definecolor{darkgreen}{rgb}{0,.5,0}
\definecolor{darkred}{rgb}{.7,0,0}

\begin{document}

\title{Gauge/Gravity Duality and the Black Hole Interior}

\author{Donald Marolf}
 \email{marolf@physics.ucsb.edu}
\affiliation{Department of Physics, University of California\\
Santa Barbara, CA 93106-9530 USA
}

\author{Joseph Polchinski}
 \email{joep@kitp.ucsb.edu}
\affiliation{Kavli Institute for Theoretical Physics, University of California \\
Santa Barbara, CA 93106-4030 USA
 }

\date{June 22, 2013}

\begin{abstract}
We present a further argument that typical black holes with field theory duals have firewalls at the horizon. This argument makes no reference to entanglement between the black hole and any distant system, and so is not evaded by identifying degrees of freedom inside the black hole with those outside.  We also address the ER=EPR conjecture of Maldacena and Susskind, arguing that the correlations in generic highly entangled states cannot be geometrized as a smooth wormhole.
\end{abstract}

\maketitle


\section{Introduction}

Gauge/gravity duality in its various forms \cite{Banks:1996vh,Maldacena:1997re} provides a construction of quantum gravity in spacetimes with special boundary conditions.  This construction is algorithmic, following from that \cite{Wilson:1993dy} of the dual field theory (DFT).  It has already provided important insights --- most notably, black holes evolve according to the rules of ordinary quantum mechanics, at least from the point of view of an external observer.   Here we apply this to examine recent claims~\cite{Almheiri:2012rt} and counterclaims regarding the nature of the black hole interior.

In \S2 we present an argument that, without other significant departures from effective field theory (EFT), typical black holes must have firewalls at their horizons.
Though this rests on assumptions similar to those of Ref.~\cite{Almheiri:2012rt} (see also \cite{Sorkin:1997ja,Mathur:2009hf,Braunstein:2009my,Giddings:2011ks}), it differs in structure, sharpening {both} an argument of Bousso~\cite{Bousso} and one from section 5 of \cite{Almheiri:2013hfa}.  In particular, these do not assume the black hole to be highly entangled with another system.
As a result, the firewall cannot be ameliorated by identifying degrees of freedom inside the black hole with those outside as suggested by \cite{Papadodimas:2012aq,Verlinde:2012cy,Maldacena:2013xja}.    Similar conclusions, based on other arguments, have appeared in Refs.~\cite{Avery:2012tf,VanRaamsdonk:2013sza}.

In \S3 we expand on commentary from Ref.~\cite{Almheiri:2013hfa} regarding potential limitations on the ability to reconstruct the interior from the \mbox{DFT}. Sec.~4 then examines the ER=EPR conjecture of Ref.~\cite{Maldacena:2013xja}.  We argue that generic entangled states of two field theories do not have a geometric interpretation in terms of a connecting wormhole.  This result may also be related to the ideas of \cite{VanRaamsdonk:2010pw}.  In \S5 we argue that the usual calculation of the Hawking flux is valid even with the firewall.

\section{Typicality of firewalls}

In the DFT, we can construct any state from the vacuum by coupling to local sources.   In the gravitational description, this corresponds to throwing excitations in from the boundary and/or allowing them to come out.  We consider an energy $E$ sufficiently high that almost all states are black holes (though recall that this can be well-below the Hawking-Page transition in AdS).  While our black holes have been explicitly formed by collapse,\footnote{It is often stated that black holes formed from collapse occupy only a subspace of small entropy.  This statement applies when the collapse occurs quickly, on timescales comparable to the AdS (anti-de Sitter) time in an AdS context.  The notion of collapse used here is broader, encompassing much slower processes that may require exponentially large times.} to high accuracy they resemble the eternal black holes that dominate any corresponding microcanonical ensemble.

Let us focus on a mode $f$ of some linearized field in this eternal black hole spacetime. We take $f$ to be a smooth wavepacket localized in Killing frequency near some positive $\omega$. We also require $f$ to strictly vanish inside and in a neighborhood of the horizon; this implies some small negative frequency tail with negligible physical effects.  In the DFT this bulk operator has an image $\hat b$ which can be obtained by expanding $\phi$ in Fourier modes and using the gauge/gravity dictionary to relate such modes to boundary operators\footnote{At high angular momenta, this construction involves large cancelations.}.  The free field expansion is valid when the appropriate parameter $N$ is infinite and can be systematically improved with $1/N$ \cite{Kabat:2011rz,Heemskerk:2012mn}, though we will not need such high precision below.

The DFT state may be expanded in eigenstates of the number operator $N_b = \hat b^\dagger \hat b$.  {EFT across the horizon relates the Hawking modes $b$ to the modes $a$ of an infalling observer by a Bogoliubov transformation such that,
assuming $N_a$ to also have some image in the DFT, the $a$-vacuum has a thermal distribution of $N_b$. } Any $N_b$ eigenstate thus differs at $O(1)$ from the $a$ vacuum, and so the expectation value of $N_a$ is at least $O(1)$ in any $N_b$ eigenstate.  In fact, {as noted in particular by Bousso~\cite{Bousso},  eigenstates of $N_b$ provide a complete basis of unentangled black hole states, all of which have firewalls.}  In the gauge/gravity context we can sharpen this to show that {\it typical} states have firewalls.  Since these basis states approximate energy eigenstates, we may use this basis  to compute the average
\begin{equation}
\overline{\! N_a \!}' = {\rm Tr}' N_a/ {\rm Tr}' 1 \,,
\label{average}
\end{equation}
where the primes indicate that we restrict to some chosen energy range.\footnote{For $N_b$-eigenstates to be a good basis, the width of this range should not be parametrically smaller than $T \sqrt{\bar N_b{}'}$, where $\bar N_b{}'$ is the corresponding average of $N_b$ and $T$ is the black hole temperature. But since large $\bar N_b{}'$ would in any case indicate a firewall, we may in fact choose any width not paramerically smaller than $T$.}${}^,$ \footnote{ If the dual theory has multiple noninteracting sectors we restrict the energy in each sector.  Thus the argument applies to the setting of Ref.~\cite{Maldacena:2013xja}.} A similar calculation in bulk EFT would be meaningless due to near-horizon divergences, but the finite density of DFT states makes \eqref{average} well-defined.

The manifest positivity of $N_a$ forbids cancelations, so our average must be at least $O(1)$.  We exclude the possibility that \eqref{average} is dominated by a small number of highly excited states, with the rest unexcited, by considering
$(1 - P_0)$ in place of $N_a$, where $P_0$ projects onto $N_a = 0$.  Since the operators for orthogonal modes approximately commute, this argument applies independently to {each} mode. {Thus we} conclude that at a significant fraction of the infalling modes are excited in a typical state: that is, most black hole states have firewalls.

Although quite different in structure from the entanglement argument of \cite{Almheiri:2012rt} (it instead resembles an argument from section 5 of \cite{Almheiri:2013hfa}), we can identify common assumptions.  Our use of the finite density of states is closely related to the unitarity assumption (postulate 1) of \cite{Almheiri:2012rt}, and both rely on bulk {EFT} in relating $b$ to {both $\hat b$ and $a$ (postulates 2 and 4).
To invalidate this argument by modifying EFT outside}, one would need {the probability distribution for $N_b$ evolve by an $O(1)$ amount} while propagating outward from the near-horizon region (as in \cite{Giddings:2011ks}). This is a large violation of effective field theory for which it is difficult to find a consistent model~\cite{Almheiri:2012rt,Almheiri:2013hfa}.

\section{Observing the interior}

If gauge/gravity duality were as complete as {one} might hope, a simple way to resolve the firewall question would be as follows: identify the field theory operator $\hat T_{\mu\nu}$ dual to the matter energy momentum tensor $T_{\mu\nu}$  behind the horizon, and calculate its expectation value in any state.  However, two severe problems present themselves.  First, there is no simple dictionary for bulk operators, and attempts to construct these seem to require that one have a complete understanding of bulk dynamics already.  Second, it is argued in Ref.~\cite{Almheiri:2013hfa} that there can be no perfect map between bulk and field theory dynamics.   In particular, $T_{\mu\nu}$ involves the Hawking partner modes $\tilde b$, and a simple counting argument showed that there is no field theory operator with the properties
\begin{equation}
[\hat {\tilde b}, \hat{\tilde b}{}^\dagger] = 1 \,,\quad [H, \hat{\tilde b}{}^\dagger] = -\omega \hat{\tilde b}{}^\dagger \,.
\label{comms}
\end{equation}
The second of these reflects the fact that the Hawking partner modes have negative ADM energy.
The interpretation of this result is not clear, and we offer here four possibilities:

\smallskip

{\it 1. Typical states have no interior.}

\smallskip

{\it 2. Typical interior states are highly excited.}  In this case there could be large corrections to the commutators~(\ref{comms}), but again there is a firewall.\footnote{Options 1 and 2 are both consistent with the viewpoint of \cite{Mathur:2013gua}. }

\smallskip
{\it 3. Strong complementarity.} The strong complementarity proposal \cite{Bousso:2012as,Harlow:2013tf} states that the theory describing an infalling observer is well approximated by effective field theory, but differs from that describing asymptotic observers (see also earlier ideas in \cite{Banks:2012nn} and \cite{VanRaamsdonk:2010pw}). Thus, in our context, it also differs from the DFT.  This might at first seem to fit well with the conventional picture of  black holes, where the adiabatic principle implies that the infalling observer sees vacuum, and not one of the many excited states.  But the argument of \S2 still applies.  For any DFT  state $|i \rangle$, measurements by the infalling observer of the modes $b$ and $\tilde b$  should be described by some density matrix $\rho^{(i)}$.  The mode $b$ is also visible to the asymptotic observer, so complementarity requires that the reduced density matrix for $b$  be the same whether obtained directly from $|i\rangle$ or from  $\rho^{(i)}$.\footnote{The same is in fact true for any operator that can be mapped to the DFT since any state may be studied by the infaller in the far past, long before she enters the black hole.  This allows an accurate analysis of its action on DFT operators, as well as accurate comparison with asymptotic observers.  The Hilbert space ${\cal H}_{SC}$ used by strong complementarity to describe the infaller's physics must thus be rather large, and in particular takes the form ${\cal H}_{DFT} \otimes {\cal H}_{sup}$. Here ${\cal H}_{sup}$ describes the action of the state on any additional observables available to the infaller. Strong complementarity then reduces to something resembling the superselection sectors of \cite{Marolf:2008tx}.  Furthermore, choosing a preferred vacuum $|0\rangle_{SC} \in {\cal H}_{SC}$ which agrees with the DFT vacuum on DFT observables identifies ${\cal H}_{DFT}$ with the subspace of ${\cal H}_{SC}$ states that form by collapse from $|0\rangle_{SC}$.  In this sense, ${\cal H}_{DFT}$ is precisely the subspace of ${\cal H}_{SC}$ singled out by the extreme cosmic censorship of \cite{Page:2013mqa}.}

\smallskip

{\it 4. State-dependence.}  Typical field theory states have a thermal spectrum for the mode $b$.\footnote{This implies that the basis states used in \S2 are not typical states, but we have used them only to evaluate the basis-independent average~(\ref{average}).}
Any such typical state can then be used to construct a representation of the commutators~(\ref{comms})~\cite{Papadodimas:2012aq,Verlinde:2012cy}.  The counting argument implies that this cannot generate the full Hilbert space of the \mbox{DFT}.  Moreover, it depends on the choice of the initial state: in order to have an interior interpretation of a given dual state, one must first
`pin the tail on the quantum donkey' \cite{HVtalk}, choosing the state on which to build the construction.  Interior observables would then be maps ${\cal H} \times {\cal H} \to {\cal H}$, outside the normal rules of quantum mechanics~\cite{Bousso:2012as,Bousso,Almheiri:2013hfa}.  This is a rather radical modification: ``God not only plays dice, She also plays pin the tail on the quantum donkey.''  It remains to be seen whether a sensible theory can be constructed on this basis.  As one challenge, any external interaction of the black hole will change the microstate, moving the system out of the original representation; thus, the base state must evolve dynamically, in addition to the usual evolution of the state of the system.  How is this to be described?

\section{Entanglement and geometry}

We now turn to dual theories highly entangled with external systems.  Ref.~\cite{Maldacena:2013xja} has made the remarkable suggestion that such entanglement is geometrized as a connecting wormhole.  This was summarized by the slogan ER=EPR, where ER refers to the wormhole as an Einstein-Rosen bridge and EPR refers to the famous Einstein-Podolsky-Rosen discussion of entanglement.  The suggestion was based largely on intuition \cite{Maldacena:2001kr} from the thermofield state.  But, {as acknowledged in \cite{Maldacena:2013xja}, the entanglements in this state are very special \cite{Louko,Marolf:2012xe,Hartman:2013qma}.} Indeed, \cite{Shenker:2013pqa} has shown that even small perturbations greatly change their nature.  Here we consider generic highly entangled states, and argue that there can be no geometric interpretation of correlations between the two field theories.

We first review features of the thermofield state.
The thermal two-point function is
\begin{equation}
{\rm Tr}(e^{-\beta H} A(t) B(t') ) = \sum_{\alpha,\beta} e^{-\beta E_\alpha + i (t-t') E_{\alpha\beta} } A_{\alpha\beta} B_{\beta\alpha} \,
\end{equation}
where $\alpha,\beta$ label energy eigenstates, $E_{\alpha\beta} = E_\alpha - E_\beta$, and $A_{\alpha\beta}$, $B_{\beta\alpha}$ are corresponding matrix elements of $A(t=0)$, $B(t'=0)$.
Continuing in $t$ gives the opposite-side correlator
\begin{eqnarray}
&&{\rm Tr}(e^{-\beta H} A(t-i\beta/2) B(t') ) \nonumber\\
\quad&& \qquad= \sum_{\alpha,\beta} e^{-\beta (E_\alpha + E_\beta)/2 + i (t-t') E_{\alpha\beta} } A_{\alpha\beta} B_{\beta\alpha}
\label{opp}
\\
&&  \qquad= \sum_{\alpha,\beta, \gamma, \delta} e^{  -i t E_{\delta\gamma} - i t' E_{\alpha\beta}} \psi^*_{\delta\beta} \psi_{\gamma\alpha}A_{\gamma\delta} B_{\beta\alpha}    \label{genopp}\\
&& \qquad\equiv \langle\psi| A_L (-t) B_R(t')| \psi\rangle\,,
\end{eqnarray}
where $\psi$ is the thermofield state
\begin{equation}
\psi_{\alpha\gamma} = Z^{-1/2} \delta_{\alpha\gamma} e^{-\beta E_\alpha/2} \,.
\end{equation}

For small $t-t'$ this expectation value is $O(1)$ in terms of the density of states $e^S$, and can be calculated geometrically in terms of the two-sided black hole \cite{Maldacena:2001kr,Hartman:2013qma}.  However, at times greater than the Page time $O(S)$, its exponentially falling geometrical value is less than its long-time average magnitude $e^{-S}$.  The latter, estimated from $e^{2S}$ terms of magnitude $e^{-2S}$ with random phases, dominates past the Page time.  Refs.~\cite{Maldacena:2001kr,Hawking:2005kf} have suggested a geometric interpretation for the long term average.  There are problems with this~\cite{Barbon:2003aq}, but at any rate it involves no geometric connection between the two boundaries.   Further, it is difficult to see how a geometric construction could be sensitive to the fine structure of the field theory spectrum.

We therefore identify correlators that are dominated by random phase behavior and suppressed by some power of the density of states $e^S$ as indications of EPR without ER, entanglement that cannot be interpreted in terms of a smooth geometric wormhole.  {For contrast, while nearly extreme Reissner-Nordstrom black holes lead to small correlators between uncharged fields, an analysis like that above shows that this is due not to erratic phases but instead to small matrix elements between the relevant energy eigenstates. The latter is natural since such eigenstates minimize the possible energy given the charge, while the operators tend to increase this energy.}

Suppose that we consider {non-thermofield} states that are equally entangled.  Is there again a geometric interpretation over some range of times, or is the correlator always nongeometric in the sense just described?
For example, suppose we have two copies of the dual theory, each beginning in some high energy pure state.  We allow them to interact for some period of time so that the pair comes to a state of thermal equilibrium. We then turn the interaction off.  Of course, if the coupling were changed sufficiently slowly, the adiabatic principle would imply that the final state is the same as the initial up to symmetry (assuming, as appropriate for a chaotic system, that there is no degeneracy of states aside from that implied by symmetry).  But we will switch the interaction on and off on some time scale $\tau$ that is long compared to typical relaxation time scales but short compared to the inverse splitting $O(e^S)$, so that the final state involves a superposition with some energy width $1/\tau$ that is narrow but contains many states.

To address this question we need to understand generic properties of the matrix elements of $A$ and $B$.  For chaotic systems we take this to be given by the {eigenstate} thermalization hypothesis (ETH)~\cite{shredder}
\begin{eqnarray}
A_{\alpha\beta} &=& {\cal A}(\bar E^{\alpha\beta} ) \delta_{\alpha\beta} + e^{- S(\bar E^{\alpha\beta})/2} f^A(E_\alpha,E_\beta) R^A_{\alpha\beta} \,,
\nonumber\\
B_{\alpha\beta} &=& {\cal B}(\bar E^{\alpha\beta}) \delta_{\alpha\beta} + e^{-S(\bar E^{\alpha\beta})/2} f^B(E_\alpha,E_\beta) R^B_{\alpha\beta} \,.
\label{eth}
\end{eqnarray}
Here, ${\cal A}$, $\cal B$, and $f^{A,B}$ are smooth functions of their arguments, and $\bar E^{\alpha\beta} = (E_\alpha + E_\beta)/2$.  The complex $R^{A,B}_{\alpha\beta}$ vary erratically, with zero mean and unit average magnitude.  They have phase correlations,
\begin{equation}
R^A_{\alpha\beta} R^B_{\gamma\delta} = \delta_{\alpha\delta} \delta_{\beta\gamma} \sigma^{AB}(E_\alpha,E_\beta) + {\rm erratic}\,,
\label{rrav}
\end{equation}
where $ \sigma^{AB}(E_\alpha,E_\beta)$ is again {smooth.}  This form is necessary for the success of statistical mechanics.  As one check, it is stable under multiplication of operators.

We now apply this to the correlator~(\ref{genopp}) for a general state.  Since we are only interested in the connected correlator, we consider operators with vanishing one-point functions, ${\cal A} = {\cal B} = 0$.  Then
\begin{eqnarray}
&&\langle\psi| A_L (-t) B_R(t')| \psi\rangle\nonumber\\
&&\quad =  \sum_{\alpha,\beta, \gamma, \delta} \Bigl\{ e^{   - i t E_{\delta\gamma} - i t' E_{\alpha\beta} } \psi^*_{\delta\beta} \psi_{\gamma\alpha} e^{-S(\bar E^{\alpha\beta})/2 - S(\bar E^{\gamma\delta})/2} \nonumber\\
&& \quad\qquad\qquad\times f^A(E_\gamma,E_\delta)f^B(E_\beta,E_\alpha) R^A_{\gamma\delta} R^B_{\beta\alpha}\Bigr\} \,.
\label{general}
\end{eqnarray}
The contribution of the smooth term in the product (\ref{rrav}) is
\begin{eqnarray}
&& \sum_{\alpha,\beta} \Bigl\{ e^{  i (t-t') E_{\alpha\beta} } \psi^*_{\beta\beta} \psi_{\alpha\alpha} e^{-S(E^{\alpha\beta})} \nonumber\\
 &&\qquad \times  f^A(E_\alpha,E_\beta)f^B(E_\beta,E_\alpha)  \sigma^{AB}(E_\alpha,E_\beta) \Bigr\} \,.
\end{eqnarray}
In the thermofield state, the wavefunction components are each of order $e^{-S/2}$, so there are order $e^{2S}$ terms each of magnitude $e^{-2S}$.  For small enough times the phases are not large and the result is order one, confirming the expected result.  {But this is special to the diagonal form of the thermofield wavefunction.}  For general states, the wavefunction components are each of order $e^{-S}$ and the sum can be no larger than $e^{-S}$, even choosing the times to cancel the phases in the wavefunction.

Considering now the erratic term in the product (\ref{rrav}), the sum~(\ref{general}) has $e^{4S}$ terms each of magnitude $e^{-3S}$.  With random phases the result is of order $e^{-S}$, but we can ask whether it is larger for special choices of the $O(e^S)$ phases $e^{i t E_\alpha}, e^{it' E_\alpha}$.  For simplicity consider the case that all terms are real.  With random signs, the total correlator $C$ is governed by a gaussian distribution of width $e^{-S}$,
\begin{equation}
P(C) \sim \exp(- C^2 e^{2S} ) \,.
\end{equation}
With $2^{ e^S}$ trials associated with $e^S$ choices of sign, we can go out on the tail to $C^2 e^{2S} \sim e^S$ or $C \sim e^{-S/2}$.  This is an improvement, but still small.  Here we have used the Gaussian approximation to the binomial distribution, but the result also follows from Hoeffding's inequality for the latter.   {Another special case arises from the phase correlation when $\alpha = \beta$, $\gamma = \delta$, but here there are $e^{2S}$ terms of magnitude $e^{-3S}$.}  We conclude that for generic entangled states, even those produced by ordinary thermal equilibration, the opposite-side correlators do not have the magnitude that would be expected from a geometric wormhole connection, and are dominated by erratic behavior from the fine structure of the spectrum.  This is reminiscent of observations in Refs.~\cite{VanRaamsdonk:2010pw,Czech}.

Our result applies to opposite-side correlators of ordinary linear operators.  A final possibility is that the choice of the operators $A, B$ depends on the state $\psi$, in a way that cancels phases.  These would then be rather complicated and special nonlocal operators, not the simple local operators that normally have geometric interpretations.  This again a nonlinear modification of the usual rules for observables, similar to that discussed in \S3, item 4.  But if state-dependence is deemed acceptable one might hope to choose the relevant observables via an action principle, such that those with large correlators are singled out.

\section{Should the calculation of the Hawking flux still hold?}

We have argued that generic states of black holes contain firewalls, contradicting the expectation from effective field theory and the adiabatic principle that the infalling observer sees vacuum.  However, the most robust derivation of the Hawking flux uses precisely the latter argument: in the $a$ vacuum, the $b$-modes have a thermal density matrix.  Thus the question has been raised a number of times: does the usual result for the Hawking flux hold?  In fact, there is a simple argument to this effect.  To begin, the argument seems strong that the thermofield state is described by the extended AdS Schwarzschild geometry with bulk quantum fields in the Hartle-Hawking vacuum~\cite{Maldacena:2001kr}, so that this state gives a thermal density matrix for $b$ and for its DFT image $\hat b$.  The {eigenstate} thermalization hypothesis then implies that the same is true to high accuracy in a generic energy eigenstate, without reference to any geometry at or behind the horizon. Note that this gives a derivation of the coarse-grained results of the Hawking calculation for $b$ that also implies corrections for the fine-grained part. This comment is similar to a crucial ingredient of fuzzball complementarity \cite{Mathur:2012zp}, though we limit its scope to observations outside the horizon.

\section{Conclusion}

It is notable that many of the arguments in this paper, and in other works on the same subject, are based on logical deduction from general principles rather than a concrete realization of quantum gravity in the bulk.  Thus, while gauge/gravity duality is a powerful tool, we believe that there is a  gap in the current understanding of quantum gravity, one that must be filled in order to move on to quantum cosmology.

\smallskip

\section*{Acknowledgements}
We thank A. Almheiri, R. Bousso, B. Chowdhury, S. Giddings, D. Harlow, M. Hotta, P. Kraus, J. Maldacena, S. Mathur, D. Page, E. Silverstein, S. Shenker, M. Srednicki, D. Stanford, J. Sully, L. Susskind, D. Turton, and M.van Raamsdonk for many related discussions.  JP was supported in part by NSF grants PHY07-57035 and PHY11-25915.  DM was supported in part by NSF grant PHY12-05500 and by FQXi grant RFP3-1008.

\end{document}